\documentclass[nonacm]{acmart}
\AtBeginDocument{%
  }
\def\shorttitle{empty}
\usepackage[T1]{fontenc}
\usepackage[utf8]{inputenc}
\usepackage[english]{babel}
\usepackage{section, caption}
\usepackage[labelformat=simple]{subcaption}
\usepackage{multirow}
\usepackage{mathtools}
\usepackage{array, caption}
\usepackage{graphicx}
\usepackage{makecell}
\usepackage{setspace}
\usepackage{hyperref}
\usepackage[nameinlink,capitalise]{cleveref}
\usepackage{booktabs}

\begin{document}
\begin{flushleft}
 {\Large
 \textbf{A Review of Cognitive Readiness, Wearable Devices, and Prospects}
 }
 \newline
 Tasnim Irtifa Chowdhury, Osaka Metropolitan University, Japan\\
 Andrew Vargo, Osaka Metropolitan University, Japan\\
 Chris Blakely, KCGI, Japan\\
 Benjamin Tag, University of New South Wales, Australia\\
 Koichi Kise, Osaka Metropolitan University, Japan
\end{flushleft}
\hfill \break
In Human-Computer Interaction (HCI) and Ubiquitous Computing, the objective of optimizing device interactions and personalizing user experiences has placed a new emphasis on accurately evaluating cognitive readiness using wearable devices. Interpreting cognitive readiness in real-world scenarios is complex due to the plethora of potential physiological measures, individual variability, and the limitations of wearable devices. In this review, we present a systematic overview of key physiological measures that can be used for an in-depth assessment of cognitive readiness. These measures can serve as proxies for detailed assessments of cognitive readiness. This review serves as a tool for assessing cognitive readiness for diverse applications, with special focus on in-the-wild research settings. In addition, due to the complexity of measurements and devices, we propose the development of robust catalog for cognitive readiness measurements.

\hfill \break
Keywords: Human Computer Interaction, Ubiquitous Computing, Wearable Devices, Health Trackers, Physiological Sensing

\section{Introduction}
Cognitive readiness is a multidimensional construct that refers to the mental preparedness of an individual or team to perform tasks under varying operational conditions \cite{morrison2002cognitive,wesensten2005cognitive,o2014teaching,sommer1993cognitive}. The demand for systems that can effectively monitor and measure cognitive readiness has surged. This surge is driven by their promising potential to facilitate improved performance in various fields \cite{sommer1993cognitive,fletcher2013evolving,erdmier2016wearable,martinho2019conceptual,perrey2022training,laarni2020promoting,friedl2018military} like healthcare \cite{wu2023screening}, education and so on \cite{crameri2021review,fletcher2013evolving,martinho2019conceptual,gasparini2023behavior}. This extended diversity is due to its influence on performance and error rates. This multifaceted construct is related to numerous domains such as attention \cite{solan2001role,baker2011attention,sarter2001cognitive,awh2001overlapping}, working memory \cite{fitzpatrick2012toddler,awh2001overlapping}, mental resilience \cite{nindl2018perspectives, fletcher2013evolving}, calculative work \cite{endsley2000situation}, reaction time \cite{wesensten2005cognitive}, errors \cite{blume2005memory}, accuracy \cite{isbell2018attentional}, reasoning \cite{erbay2013predictive,bierman2009behavioral}, strategic decision-making \cite{grier2012military,morrison2002cognitive}, and distress tolerance \cite{grier2012military,safa2014effectiveness}. Therefore, accurately measuring and understanding cognitive readiness is of paramount importance in the field of cognitive science \cite{grier2012military,fiore2010toward}.

Traditionally, subjective measures like self-reports, expert observations, and cognitive task analysis were used to assess cognitive readiness \cite{paas2003cognitive,bell2008active}. In self-reports participants report about their cognitive states by themselves. Then, in expert observations, experts observe the cognitive states of participants while doing tasks. Finally, in cognitive task analysis, cognitive tasks are broken down into constituent parts, allowing for the identification and assessment of the cognitive processes required to perform each part~\cite{clark2008cognitive,clark1996cognitive}. However, all these methods often fail to provide objective data and cannot give real-time insight as they are based on observation, interview and non-real-time data. Also, the result interpretations can be biased, and thus can be subjective. However, various other methods are also used for this assessment, including mental models and simulations. In mental models, the individual's or team's understanding of the task and the environment are assessed \cite{stout1999planning}. Simulations provide a realistic environment in which cognitive readiness can be evaluated \cite{crichton2000training}. More recent methods involve physiological measurements, which promise to provide more objective measures of cognitive and mental states, such as cognitive load \cite{kort2001affective}, alertness \cite{tag2019continuous}, attention, mental resilience, decision-making and so on. Interestingly, the introduction of wearable computing devices has led to several advancements in the field of cognitive readiness analysis. By gathering the real-time physiological data in-the-wild \cite{cosoli2022accuracy}, it can provide an on-going window into an individual's cognitive readiness \cite{mandrick2016neural}. Here, the term in-the-wild means collecting the physiological data in natural, everyday settings rather than controlled laboratory environments \cite{rogers2017research,callon2003research}. In turn, this advancement can be used for the development of systems which can lower unnecessary cognitive demand and foster an environment conducive to better performance and readiness \cite{brown1999human}. However, utilizing wearables in this context is not without its challenges. First, there is the issue of ensuring compatibility across diverse populations. Physiological responses can vary significantly among individuals based on factors such as age, gender, fitness level, and cultural background, which may affect the accuracy of cognitive readiness assessments \cite{intille2007technological,betancourt1993study}. Second, the operational environment where the device is used can influence the quality and reliability of the data collected \cite{bulling2014tutorial}. Third, user comfort and adherence to wearing the device consistently pose significant challenges to generating reliable assessment results \cite{gao2015empirical,patel2015wearable}. While wearable devices can measure physiological responses, interpreting these responses in terms of specific cognitive readiness components remains a complex task \cite{melzner2014mobile,fairclough2014advances}. Finally, there are issues concerning data privacy and security, given the sensitive nature of the physiological data being collected \cite{zheng2018blockchain}. 

Despite these challenges, the use of wearable devices for monitoring cognitive readiness provides valuable data that could inform the design of intervention strategies that enhance cognitive performance, for example, developing individualized learning and training programs \cite{chan2012smart}. Therefore, this review seeks to provide a comprehensive understanding of the physiological measures linked to cognitive readiness and the potential of wearable devices to capture these measurements in the context of in-the-wild studies. We aim to outline the different physiological parameters measured and their relationship with cognitive readiness. We also discuss the various devices used to achieve these measurements across diverse populations and environments. Through this endeavor, we summarize the current state of research in this rapidly evolving field for the benefit of Human-Computer Interaction (HCI) researchers looking to run studies in-the-wild. 

\section{Background and Methodology}
This work is based on the premise that it is necessary for experimenters to measure cognitive readiness for HCI applications {for use in in-the-wild settings}. The emphasis here is not on uncovering new physiological signals that could be tied to measuring readiness but rather on what signals are known to be useful for measuring cognitive readiness and what types of wearable devices can adequately measure them {outside of laboratory situations}. To do this, we performed a scoping review of HCI applications concerned with cognitive sensing. 

In-the-wild studies form a vital part of the HCI research community, as they give insight into natural behavior and thus can work towards the creation of applications and methods which can be applied in the real world \cite{chamberlain2012research,brown2011into}. In comparison to laboratory studies, in-the-wild studies are conducted in the real world, and often focus on allowing participants to interact or use a device within the constraints of their daily lives. While this is beneficial for gathering naturalistic data in comparison to laboratory studies, there are increased costs and uncertainty with regards to data labeling \cite{vargo2021obtaining} and participant ability to use the device correctly in their daily life \cite{nolasco2023examining}. Thus, it is vital for researchers conducting in-the-wild experiments to have a clear understanding of the benefits and limitations of devices and measuring techniques that can be applied in their studies.

There have been prior review studies on cognitive readiness. Charles et al. took six physiological measures (heart, respiration, ocular, brain, blood pressure and dermal) into consideration \cite{charles2019measuring} to assess mental workload. Longo et al. \cite{longo2022human} discuss different categories of physiological and neurophysiological measures for assessing cognitive states. The measures included heart rate, heart rate variability, blood pressure, respiratory measures, ocular measures, and neuroendocrine measures. These reviews note the variety of different measures used in isolation in past research. Furthermore, they recommend future research and practices to utilize these measures together for more robust assessments. 

\section{Cardiovascular Measurements}
\subsection{Heart Rate (HR)}
HR describes the number of times the heart beats per minute. It is a crucial indicator of an individual's cardiovascular health and can respond to various physiological and psychological states \cite{mather2018heart}. 

\subsubsection{Relationship of HR with cognitive readiness}
\label{hr}
As an autonomic response to stress, HR can be used to assess cognitive readiness. It plays an important role in improving emotion regulation \cite{mather2018heart}. It has been associated with cognitive performance, particularly in tasks requiring attention and working memory \cite{thayer2009heart,thayer2009claude,veltman1996physiological,backs1994metabolic}. Under stress, the sympathetic nervous system becomes more active, leading to an increase in HR \cite{ulrich2009neural}. This increase can also be associated with the level of cognitive demand required by a particular task. Increased cognitive load is associated with lower reasoning performance scores \cite{solhjoo2019heart}. For example, the two main categories of cognitive effort considered by Fairclough and Mulder~\cite{fairclough2012psychophysiological} are computational effort, in response to increased cognitive load, and compensatory effort, used to maintain performance when fatigued. When computational effort increases due to a more challenging task, there is a rise in HR as HRV decreases, causing greater regularity in HR. This is an indication of increased cognitive load. Increased cognitive load is associated with poor performance; that is, more error and less accuracy. Increases in HR are associated with poor performance \cite{thayer2009heart}.  

In contrast, when compensatory effort increases, i.e., attention is required due to a non-optimal physiological state such as fatigue or stress, cardiovascular changes are less predictable \cite{kramer2020physiological} and heavily dependent on the specific situation. The baroreflex, a mechanism that controls short-term blood pressure, plays a significant role in determining the individual's response in these situations. Autonomic activation, triggered by mental states, influences HR and other sub-systems that manage short-term blood pressure control like the force of heart contractions, peripheral resistance of arteries, and venous return. All these sub-systems coordinate to control heart function and blood pressure levels, thus providing adequate energy to the brain depending on the current mental state. During mentally demanding tasks, HR generally increases and becomes more regular compared to resting states or easier tasks \cite{fairclough2012psychophysiological}. In some other studies, it is shown that increased HR has been associated with better cognitive performance in tasks requiring sustained attention \cite{chen2010detecting}. Deceleration of HR can also be directly correlated with reaction time \cite{obrist1969heart}. Oldehinkel et al. \cite{oldehinkel2008low} found that lower HR is a marker of mental resilience. Linden and Wolfgang \cite{linden1991arithmetic} established a relation between arithmetic (i.e., calculative) work and HR. Higher resting HR is also associated with greater cognitive distress \cite{krause2016spiritual}. Crone et al. examined HR patterns of participants engaged in the Iowa Gambling Task (IGT) and found that anticipatory HR slows down when choosing disadvantageous options \cite{crone2004heart}. Thus, HR can serve as a physiological marker for attention and working memory-related cognitive readiness.

\subsection{Heart Rate Variability (HRV)}
The amount of time between consecutive heartbeats and how it changes is known as HRV. It can provide insights into cognitive readiness by reflecting the balance between sympathetic and parasympathetic nervous system activity \cite{Thayer2012}. It has been related to activity in the prefrontal cortex, a part of the brain involved in planning complex cognitive behavior and decision-making \cite{Kim2018}. At the same time, activity in the prefrontal cortex is inversely related to activity in sub-cortical structures such as the amygdala, a region of the brain involved in emotional responses. Lane et al. \cite{Lane2017} propose that HRV can indicate activity in prefrontal neural structures and is therefore related to cognitive performance. There are several HRV indicators  \cite{allen2007many} which include time-based, frequency-based, and shape-based methods. The standard deviation of normal-to-normal (NN) intervals (SDNN), the root mean squared successive differences in intervals (RMSSD), and the proportion of interval differences exceeding 50 ms (p50) fall into the time domain of the overall variability category. The frequency domain category consists of measures like the fast Fourier transform power spectrum density in the high-frequency band, low frequency (LF), high frequency (HF), and the ratio of LF and HF (LF/HF). Finally, the HRV triangular index (HRVTI) and the baseline width of the respiratory rate histogram by triangular interpolation (TINN) are measured based on the shape of the respiration rate interval histogram \cite{grossmann2016heart}. We have considered only SDNN, RMSSD, HF, LF, and LF/HF because of their wide uses with wearable devices in cognitive analysis \cite{held2021heart,hilgarter2021phasic,sanches2023wearable,koeneman2021wearable}. 

\subsubsection{Relationship of HRV with Cognitive Readiness}
\label{hrv}
Executive functions are central to our ability to plan and direct action and thought for goal-directed behavior. These functions are seen as a central monitoring system, including aspects such as the selection, maintenance, updating, and rerouting of information. They also involve the ability to suppress irrelevant or interfering information. Sustained attention and working memory are considered core components of executive functions. Thayer et al. \cite{thayer2009heart} discuss the relationship between vagally-mediated HRV and these core components of executive function in military personnel. Vagally-mediated HRV is generally assessed with RMSSD and HF power. High-HRV individuals, those who exhibit a higher degree of variability between inter-bit interval, performed better on cognitive tasks related to attention and memory, particularly on tasks requiring executive functioning. High HRV is associated with better cognitive flexibility and adaptability under stress, which is crucial for attention, working memory, mental resilience, and distress tolerance \cite{luque2013cognitive,forte2019heart,hansen2003vagal}. Studies have also shown that HRV {is also related to individual performance and is thus an indicator for error and accuracy} \cite{thayer2009heart}. Calculative work, i.e., mental workload, decreases the power of both low frequency (0.04-0.15Hz) and high frequency (0.20-0.50Hz) components of HRV \cite{yoshino2005causal}. Porges in his study \cite{porges1972heart} finds that HRV measures are strongly correlated with reaction time. HRV also affects emotion regulation in brain networks \cite{mather2018heart} and reasoning \cite{solhjoo2019heart}. Forte et al. discuss the relationship between decision-making and HRV \cite{forte2022decision}.

\subsection{Blood Flow Rate (BFR) and Blood Oxygen Level (BOL)}
The volume of blood that flows through a vessel, an organ, or the entire circulatory system in a given period is known as BFR. The rate can change based on various factors such as physical activity, temperature, hydration, stress, and illness. During cognitive tasks, there is increased neural activity in specific regions of the brain, which results in an increase in blood flow to these areas \cite{attwell2010glial}. This relationship between local neural activity and changes in cerebral blood flow is called Neurovascular Coupling. The brain consumes a significant amount of the body's total oxygen to maintain its high Neurovascular activity. Blood flow to the brain delivers the necessary oxygen and nutrients \cite{iadecola2017neurovascular}. Therefore, a higher BFR might indicate that the brain is well-supplied with oxygen, which is essential for optimal cognitive functioning \cite{iadecola2004neurovascular,rypma1999roles}. On the other hand, conditions that limit blood flow, such as hypertension and atherosclerosis, have been associated with cognitive impairment \cite{iadecola2013pathobiology}. BOL is the measure of oxygenation levels within the blood at a given time. BOL is a similar measure to BFR as it can show cognitive performance as affected by oxygen delivery to the brain.

\subsubsection{Relationship of BFR and BOL with Cognitive Readiness}
\label{bfr}
BFR, specifically in the brain, can provide insights into cognitive readiness by revealing the allocation of neural resources. Increased cerebral blood flow has been associated with better cognitive performance in tasks requiring attention, working memory, and reasoning \cite{parkes2004normal,iadecola2004neurovascular}. BFR affects mental resilience in the brain \cite{hughes2012roots}. In calculative work, task difficulty is reflected by the change in BFR in the brain \cite{vingerhoets2002reliability}. There also exists a negative correlation between BFR and reaction time \cite{duschek2004cognitive}. Changes in BFR are an important indicator of cognitive components like decision-making \cite{osorio2015structured}, distress tolerance \cite{mariano2015effects} and emotion regulation \cite{thayer2000model}. This suggests that measuring changes in cerebral blood flow could provide insights into an individual's cognitive readiness. 
With regards to BOL, hypoxia can impair cognitive performance in tasks requiring attention, working memory, and reasoning \cite{moore2018effects}. Moreover, cognitive performance can be optimized when BOLs are maintained within an optimal range \cite{rajguru2013military}.

\section{Neurooculomuscular Measurements}
\subsection{Sleep}
Sleep is a natural and essential state that involves a periodic suspension of consciousness in which the body can rest. In this state, the subject experiences altered consciousness alongside inhibited muscular and sensory activity. It's a crucial part of our health as it allows the body to repair itself and be fit and ready for another day \cite{Durmer2005}. Moreover, it plays a vital role in brain function, including learning, memory consolidation, motor skills, and neurocognitive performance \cite{walker2009year}. The specific stages of Non-Rapid Eye Movement (NREM), Rapid Eye Movement (REM), and slow-wave sleep (SWS) \cite{walker2009year} play pivotal roles in the above activities. Sleep deprivation negatively impacts memory encoding, particularly positive memory, though not negative recall as much \cite{9115246}. Sleep deprivation can impair alertness, attention, and vigilance, leading to reduced performance and potentially serious health conditions \cite{Goel2010,Banks2010}. Technological advancements now allow for non-invasive monitoring of these rhythms and their impact on cognition. 

\subsubsection{Relationship of Sleep with Cognitive Readiness}
\label{sleep}
In terms of neurocognitive performance, restricted sleep causes performance deterioration. People with restricted sleep are more prone to error \cite{johnson2014sleep} and therefore have less accuracy. Worse sleep quality is also associated with lower mental resilience \cite{novak2023resilience}. However, "sleep banking," or increasing sleep hours before sleep restriction, can reduce these adverse effects. Extended sleeping hours can significantly reduce reaction time \cite{swinbourne2018effects}. Scarpelli et al. in their study found a relation between sleep and attention \cite{scarpelli2019advances}. Sleep quality also affects the reasoning process \cite{quevedo2023associations}. Different sleep stages have been found to play specific roles in cognitive readiness and overall cognitive performance. NREM sleep is essential for memory consolidation and learning \cite{siegel2001rem}. SWS has been associated with the restoration of cognitive function \cite{vargo2022sleep}, synaptic homeostasis, and the integration of newly acquired information. REM sleep has been linked to emotional regulation, working memory, and problem-solving, especially calculative work and decision making \cite{prinz1982changes,stickgold2013sleep}. It plays a role in distress tolerance \cite{reitzel2017distress}, the regulation of emotional experiences and the consolidation of procedural and spatial memory \cite{palmer2017sleep}.

\section{Electrodermal Measurements}
\subsection{Skin Conductance (SC)}
SC is a measure of the skin's ability to conduct electricity. It changes because of changing varying levels of moisture exuded from the skin. It is linked to the activity of the eccrine sweat glands, which are controlled by the sympathetic branch of the autonomic nervous system (ANS) \cite{boucsein2012electrodermal}. The activity of these glands, and thus SC, reflects arousal of the sympathetic ANS, a system that often accompanies various psychological processes \cite{critchley2002electrodermal}. Sweating occurs from these glands. The central control of the ANS and, consequently, eccrine sweating is complex, originating within the hypothalamus and the brainstem. However, other brain areas, like the amygdala, hippocampus, basal ganglia, and prefrontal cortex, are also involved in this control \cite{dawson2017electrodermal}. 

\subsubsection{Relationship of SC with Cognitive Readiness}
\label{sc}
These areas, part of the limbic and paralimbic networks, play a critical role in affective processes, such as emotions. Sweating rate, as measured by galvanic skin response (GSR), also known as Electrodermal Activity (EDA)~\cite{babaei2021critique}, can be influenced by emotional arousal and stress, both of which can impact cognitive readiness. Higher GSR levels have been associated with increased cognitive load, as well as emotional arousal, such as anxiety \cite{boucsein2012electrodermal,figner2011using}. Moreover, the sweating rate has been found to correlate with cognitive performance in high-stress situations \cite{saus2006effect}, suggesting that GSR might be a useful indicator of mental resilience and distress tolerance \cite{perala2007galvanic,dawson2017electrodermal,critchley2000neural}. Frith et al. \cite{frith1983skin}, engaged the subjects in activities for testing reaction time, vigilance, or calculative work while listening to irrelevant tones. SC was greater in response to these tones during tasks, indicating increased attention. SC is also an important indicator to working memory and decision-making \cite{patterson2002task}. During logical reasoning tasks, SC responses also increase \cite{spiess2007skin}. In addition to SC, elevated Skin Temperature (ST) is also an important indicator of cognitive load. It is often associated with increased stress levels \cite{havenith2005temperature}. Monitoring skin temperature changes can, therefore, help in assessing distress tolerance.

\section{Respiratory Measurements}
\subsection{Breathing Rate (BR)}
The process of inhalation and exhalation is called breathing. The BR refers to the number of breaths a person takes per minute. The normal range for an adult at rest can vary but is typically between 12 to 20 breaths per minute \cite{prause2023mechanical}. It is more than just a means of oxygenating blood and removing carbon dioxide. Studies suggest that it also plays a significant role in cognitive functioning \cite{zelano2016nasal}, such as emotion regulation \cite{mather2018heart}. BR, which can be influenced by various emotional and cognitive states, has been found to synchronize with local field potential responses in the olfactory system \cite{tang2015neuroscience}. This synchronization may help regulate cortical excitability and synchronize brain cell activity, ultimately shaping behavior and memory. Further, breathing has been found to influence high-frequency oscillations in both olfactory and non-olfactory brain regions, potentially affecting a variety of behaviors and learning \cite{zelano2016nasal}. Controlled breathing is also found to affect sympathetic and parasympathetic autonomic activity. It can also enhance mental performance \cite{deepeshwar2022slow}. In several studies, reduced BR or breath retention is analyzed as a measure of distress tolerance \cite{sutterlin2013breath,kraemer2016mind}. BR also has a direct correlation with attention \cite{mitsea2022breathing}. Calculative works are also found to induce changes in BR \cite{fokkema2006different}. 

\subsubsection{Relationship of BR with Cognitive Readiness}
\label{br}
Although further research is required to understand the relationship between BR and cognitive readiness, specifically how respiratory rhythms may influence cortical oscillations in the human brain, BR can influence cognitive performance through its impact on oxygen delivery to the brain and modulation of arousal. Studies have shown that slow, deep breathing can improve cognitive performance in tasks requiring attention and working memory by increasing oxygen delivery to the brain and reducing arousal levels \cite{zelano2016nasal}. BR has also been associated with mental resilience, as it can be voluntarily controlled to modulate stress responses and improve cognitive flexibility \cite{ma2017effect,tang2015neuroscience}. Deepeshwar et al. suggested that slow yoga breathing practice (SYB) at a fixed breathing rate has a direct influence on working memory \cite{deepeshwar2022slow}. Gallego et al. suggested that reaction time is longer during SYB than spontaneous breathing \cite{gallego1991assessing}. Controlled BR is also found to have a positive effect on better decision-making \cite{de2019breathing}. The potential relationship between BR, reaction time, errors, and accuracy requires further exploration. 

\section{Ocular Measurements}
\subsection{Pupil Dilation (PD)}
PD is the expansion of the pupils in response to various stimuli, such as changes in light, arousal, cognitive processing, or the use of certain medications or substances. Pupillometry is a process in which dilation of the pupils is measured and analyzed \cite{laeng2012pupillometry}. This is often used as a physiological indicator of cognitive load or mental effort, which can subsequently provide insight into a person's cognitive readiness. 

\subsubsection{Relationship of PD with Cognitive Readiness}
\label{pd}
According to the cognitive load theory, the size of our pupils changes with varying degrees of cognitive load. This theory was first developed by Sweller (1988) \cite{sweller1988cognitive} and is built on the assumption that our cognitive resources are limited, so when we're engaging in a difficult task, more cognitive resources are devoted to that task, which may be observed through larger pupil dilation. Pupil dilation can occur in response to a range of cognitive tasks, including perception, attention, working memory, and calculative work \cite{beatty2000handbook}. This phenomenon is known as the Task-Evoked Pupillary Response (TEPR). Cognitive effort and cognitive capacity are intrinsically linked to PD as stated by the dual-process theory \cite{kahneman1973attention}. According to this theory, when cognitive capacity is exceeded, cognitive readiness decreases and this may be reflected in greater PD. PD has a clear dissociation with reaction time \cite{hershman2019dissociation}. Research using functional Magnetic Resonance Imaging (fMRI) has shown that changes in the locus coeruleus, a part of the brainstem that responds to stress, are correlated with pupil diameter and are involved in regulating attention and behavior in a person \cite{aston2005integrative}. Increased PD has been observed in tasks requiring attention, working memory, and mental resilience \cite{kahneman1966pupil,peavler1974pupil}. Whereas, decreased PD is associated with metacognitive accuracy \cite{van2018pupil,lempert2015relating}. Additionally, PD can be used to infer cognitive flexibility, as individuals who are more cognitively flexible tend to exhibit larger pupil dilation in response to distress \cite{van2010resource,prehn2011influence}. In analogical reasoning tasks, PD has proven itself as a useful marker both in decision-making and emotion regulation \cite{prehn2011influence,prehn2013pupil,tsukahara2016relationship}.

\subsection{Eyeblink (EB)} 
EB is a semi-voluntary, rapid closing of the eyelid. A blink is controlled by two muscles: the orbicularis oculi and levator palpebrae superioris. It is a complex neuromuscular response. Various stimuli trigger this response. Environmental factors such as wind, foreign bodies, or bright lights can be triggers. Psychological factors like stress or cognitive workload, as well as social context~\cite{gupta2019blink}, can also affect eye blinking \cite{stern1984endogenous}.

\begin{table*}[!htbp]
    \centering
 \captionof{table}{{Overview of cognitive readiness assessed with different physiological measures}}
  \begin{tabular}{lll||lll}
        \toprule
         \multirow{2}{4em}{Cognitive Readiness} & \multirow{2}{6em}{Physiological parameters} & {References} &\multirow{2}{4em}{Cognitive Readiness} & \multirow{2}{6em}{Physiological parameters} & {References}   \\ 
         & & & & & \\
         \midrule
         Attention & HR  (\ref{hr})& \cite{thayer2009heart,thayer2009claude,veltman1996physiological,backs1994metabolic} & Working & HR  (\ref{hr})& \cite{thayer2009heart,veltman1996physiological,backs1994metabolic}\\
         & HRV  (\ref{hrv})& \cite{thayer2009heart} & Memory & HRV  (\ref{hrv})& \cite{thayer2009heart}\\
         & BFR \& BOL  (\ref{bfr})& \cite{parkes2004normal} & & BFR \& BOL  (\ref{bfr})& \cite{iadecola2004neurovascular}\\
         & Sleep  (\ref{sleep})& \cite{scarpelli2019advances}& & Sleep  (\ref{sleep})& \cite{vargo2022sleep}\\
         & SC  (\ref{sc})& \cite{frith1983skin}& & SC  (\ref{sc})& \cite{patterson2002task}\\
         & BR  (\ref{br}) & \cite{mitsea2022breathing} & & BR  (\ref{br}) & \cite{backs1994metabolic,zelano2016nasal}\\
         & PD  (\ref{pd})& \cite{beatty2000handbook,kahneman1966pupil, peavler1974pupil}& & PD  (\ref{pd})& \cite{beatty2000handbook,kahneman1966pupil, peavler1974pupil}\\
         & EB  (\ref{eb})& \cite{jongkees2016spontaneous,maffei2018spontaneous,sakai2017eda} & & EB  (\ref{eb})&\cite{jongkees2016spontaneous, ortega2022spontaneous}\\
          & Posture  (\ref{posture})& \cite{smith2019standing,langton2000eyes,mehrabian1968relationship} & & Posture (\ref{posture})& \cite{peper2012increase, riley2003inverse}\\
         & PA  (\ref{pa})& \cite{kramer2006exercise,vanhelst2016physical}& & PA  (\ref{pa})&\cite{hillman2008smart}\\
         & BW  (\ref{bw})& \cite{chen2017assessing,thomas2016eeg}& & BW  (\ref{bw})& \cite{roux2014working,ruchkin1995working}\\
         & Hydration  (\ref{hydration})& \cite{gopinathan1988role,cian2001effects, patel2007neuropsychological}& & Hydration  (\ref{hydration})& \cite{gopinathan1988role,cian2001effects,patel2007neuropsychological}\\ [.30cm] 
         Mental  & HR  (\ref{hr})& \cite{oldehinkel2008low} & Calculative & HR  (\ref{hr})& \cite{fairclough2012psychophysiological,linden1991arithmetic,schuri1981heart}\\
         Resilience & HRV  (\ref{hrv})& \cite{thayer2009heart,luque2013cognitive} & Work & HRV  (\ref{hrv})& \cite{yoshino2005causal}\\
         & BFR \& BOL  (\ref{bfr})& \cite{hughes2012roots} & & BFR \& BOL  (\ref{bfr})& \cite{vingerhoets2002reliability}\\
         & Sleep  (\ref{sleep})& \cite{novak2023resilience} & & Sleep  (\ref{sleep})& \cite{vargo2022sleep}\\
         & SC  (\ref{sc})& \cite{perala2007galvanic}& & SC  (\ref{sc})& \cite{frith1983skin}\\
         & BR  (\ref{br})& \cite{ma2017effect}& & BR  (\ref{br})& \cite{fokkema2006different}\\
         & PD  (\ref{pd})& \cite{peavler1974pupil,kahneman1966pupil}& & PD  (\ref{pd})& \cite{beatty2000handbook}\\
         & Posture  (\ref{posture})& \cite{wilson2004effects,nair2015slumped} & & EB  (\ref{eb})& \cite{paprocki2017does,schuri1981heart}\\
         & PA  (\ref{pa})& \cite{belcher2021roles, hegberg2015physical}& & Posture  (\ref{posture})& \cite{riley2003inverse, kuroishi2014deficits}\\
         & BW  (\ref{bw})& \cite{davidson2004does, uhlhaas2017youth}& & PA  (\ref{pa})& \cite{have2018classroom,pindus2016moderate}\\
         & & & & BW  (\ref{bw})& \cite{pauli1994brain,crk2015understanding}\\
         & & & & Hydration  (\ref{hydration})& \cite{gopinathan1988role,cian2001effects,patel2007neuropsychological}\\ [.30cm]
          Reaction & HR  (\ref{hr})& \cite{obrist1969heart} & Accuracy & HR  (\ref{hr})& \cite{thayer2009heart}\\
         Time & HRV  (\ref{hrv})& \cite{porges1972heart}& & HRV  (\ref{hrv})& \cite{thayer2009heart}\\
         & BFR \& BOL  (\ref{bfr})& \cite{duschek2004cognitive}& & Sleep  (\ref{sleep})& \cite{johnson2014sleep} \\
         & Sleep  (\ref{sleep})& \cite{swinbourne2018effects}& & PD  (\ref{pd})& \cite{van2018pupil,lempert2015relating}\\
         & SC  (\ref{sc})& \cite{frith1983skin} & &  EB  (\ref{eb})&\cite{ortega2022spontaneous}\\
         & BR  (\ref{br})&\cite{gallego1991assessing} & & PA  (\ref{pa})& \cite{caponnetto2021effects,booth2013associations}\\
         
         & EB  (\ref{eb})& \cite{fogarty1989eye,oh2012spontaneous}& & BW  (\ref{bw})& \cite{falkenstein1991effects,craig2012regional}\\
         & PA  (\ref{pa})& \cite{brisswalter1997influence,abourezk1995effect}& & Hydration  (\ref{hydration})& \cite{gopinathan1988role,cian2001effects,patel2007neuropsychological}\\
         & Posture  (\ref{posture})& \cite{vuillerme2000effects,nashner1981relation}& & & \\
         & BW  (\ref{bw})& \cite{luck2005introduction,binias2023analysis} & &\\
         & Hydration  (\ref{hydration})& \cite{gopinathan1988role,cian2001effects,patel2007neuropsychological}& & \\ [.30cm]
         Reasoning & HR  (\ref{hr})& \cite{solhjoo2019heart} & Decision & HR  (\ref{hr})& \cite{crone2004heart}\\
          & HRV (\ref{hrv}) &\cite{solhjoo2019heart} & Making& HRV  (\ref{hrv})& \cite{forte2022decision}\\
         & BFR \& BOL  (\ref{bfr})& \cite{iadecola2017neurovascular} & & BFR \& BOL  (\ref{bfr})& \cite{osorio2015structured}\\
         & Sleep  (\ref{sleep})& \cite{quevedo2023associations} & & Sleep  (\ref{sleep})& \cite{prinz1982changes}\\
         & SC  (\ref{sc})& \cite{spiess2007skin}& & SC  (\ref{sc})& \cite{patterson2002task}\\
         & PD  (\ref{pd})& \cite{prehn2011influence,prehn2013pupil,tsukahara2016relationship}  & & BR  (\ref{br})& \cite{de2019breathing}\\
         & PA  (\ref{pa})& \cite{perrot2009physical, wandzilak1988values} & & PD  (\ref{pd})& \cite{prehn2011influence,prehn2013pupil}\\
        & BW  (\ref{bw})& \cite{bonnefond2012alpha, kao2012analysis}   &  & EB  (\ref{eb})& \cite{barkley2017eye}\\
         & Hydration  (\ref{hydration})& \cite{lieberman2007hydration,del2012hydration}& & Posture (\ref{posture})& \cite{liu2021stand}\\
         &  & & & PA  (\ref{pa})& \cite{tomporowski2003effects}\\
          &  & & & BW  (\ref{bw})& \cite{sanfey2003neural,craig2012regional}\\
         & & & & Hydration  (\ref{hydration})& \cite{del2012hydration}\\ [.30cm]
            \bottomrule
    \end{tabular}
        \label{table1}
\end{table*}

\begin{table*}[!t]
    \centering
 \captionof{table}{{Overview of cognitive readiness assessed with different physiological measures (contintued)}}
  \begin{tabular}{lll||lll}
        \toprule
         \multirow{2}{4em}{Cognitive Readiness} & \multirow{2}{6em}{Physiological parameters} & {References} &\multirow{2}{4em}{Cognitive Readiness} & \multirow{2}{6em}{Physiological parameters} & {References}   \\ 
         & & & & & \\
         \midrule
 Distress & HR  (\ref{hr})& \cite{ulrich2009neural, krause2016spiritual}& Emotion & HR  (\ref{hr})& \cite{mather2018heart}\\
         & HRV  (\ref{hrv})& \cite{hansen2003vagal} & Regulation & HRV  (\ref{hrv})& \cite{mather2018heart}\\
         & BFR \& BOL  (\ref{bfr})& \cite{mariano2015effects}& & BFR \& BOL  (\ref{bfr}) & \cite{thayer2000model}\\
         & Sleep  (\ref{sleep})& \cite{reitzel2017distress}& & Sleep  (\ref{sleep})& \cite{palmer2017sleep}\\
         & SC  (\ref{sc}) & \cite{critchley2000neural}& & SC  (\ref{sc})&\cite{boucsein2012electrodermal} \\
         & BR  (\ref{br})& \cite{kraemer2016mind}& & BR  (\ref{br})& \cite{mather2018heart}\\
         & PD  (\ref{pd})&\cite{van2010resource,prehn2011influence} & & PD  (\ref{pd})& \cite{prehn2011influence,prehn2013pupil}\\
         & EB  (\ref{eb})& \cite{maffei2019spontaneous} & & EB  (\ref{eb})& \cite{jackson2003now,jackson2000suppression}\\
         & Posture  (\ref{posture})& \cite{comaru2009postural,franklin2018physical}& & Posture  (\ref{posture})& \cite{riskind1982physical,reed2020body}\\
         & PA  (\ref{pa})& \cite{sciamanna2017physical} & & PA  (\ref{pa})& \cite{tang2022relationship,aparicio2016role}\\
         & BW  (\ref{bw})& \cite{flo2011transient}& & BW  (\ref{bw})& \cite{park2022effects,pollatos2012attenuated}\\
         & & & & Hydration  (\ref{hydration})& \cite{del2012hydration}\\ 
          \bottomrule
          \end{tabular}
        \label{table2}
\end{table*}

\subsubsection{Relationship of EB with Cognitive Readiness}
\label{eb}
The relationship between eye blinking and cognitive readiness or mental workload is quite intricate. Research suggests that spontaneous eye blink rate can be an indirect indicator of dopamine function in the central nervous system, which plays a role in regulating various cognitive functions, including attention \cite{maffei2018spontaneous} and working memory  {related to task accuracy} \cite{jongkees2016spontaneous,ortega2022spontaneous}. One theory \cite{stern1994blink} suggests that when an individual is more cognitively loaded or engaged in a task, the blink rate tends to decrease. This decrease in blink rate might be due to the need to gather more visual information and maintain visual attention during the cognitive task. Paprocki et al. \cite{paprocki2017does} suggests that blink rate is related to calculative work - the higher the blink rate, the more accurately a person answers calculative tasks. The blink rate is task-dependent where more reaction time is needed to blink during analytical tasks than during detective tasks \cite{fogarty1989eye}. Jackson et al. found that suppression of a negative emotion tends to decrease EB \cite{jackson2003now,jackson2000suppression}.

\section{Musculoskeletal Measurements}

\subsection{Posture}
Posture refers to how someone holds their body when standing, sitting, or lying down. It is the alignment and arrangement of body parts in relation to each other and in relation to the environment. Good posture involves the positioning of the joints in a manner that minimizes stress and strain on the muscles and ligaments while performing activities or at rest \cite{mcgill2015low}.

\subsubsection{Relationship of Posture with Cognitive Readiness}
\label{posture}
Posture can influence cognitive readiness by affecting physical comfort, arousal, and cognitive resource allocation. An individual's physical orientation can be a sign of their attention focus. This is rooted in the idea of joint attention, the shared focus of two individuals on an object \cite{langton2000eyes}. The degree to which a person leans towards or away from something can indicate interest or attention. Studies found that individuals tend to lean towards those they like and away from those they dislike \cite{mehrabian1968relationship}. Head and eye position can provide a nonverbal cue of someone's attentional focus. People tend to fixate their gaze on what they're attending to \cite{yarbus1967eye}. On the other hand, excessive movement or fidgeting may be a sign of restlessness or lack of focus \cite{sarver2015hyperactivity}. Upright posture has been associated with improved cognitive performance in tasks requiring attention and working memory, as it promotes better oxygenation and blood flow to the brain \cite{peper2012increase}. Additionally, maintaining an upright posture can enhance mental resilience by facilitating the regulation of stress responses \cite{wilson2004effects,nair2015slumped}. Smith et al. found that a person standing has more attention than when sitting \cite{smith2019standing}. While standing, working memory and calculative work are seen to be associated with enhanced postural stability. Postural sway reduces while doing difficult calculative work \cite{riley2003inverse,kuroishi2014deficits}. Foot pressure is significantly decreased while performing reaction-related tasks \cite{vuillerme2000effects,nashner1981relation}. Posture also has effect on decision-making \cite{liu2021stand}, distress tolerance \cite{comaru2009postural,franklin2018physical} and emotion regulation \cite{riskind1982physical,reed2020body}.

\subsection{Physical Activity (PA)}
Any movement of the body that is produced by skeletal muscles and requires energy expenditure is called PA. This includes activities undertaken while working, playing, carrying out household chores, traveling, and engaging in recreational pursuits. It is fundamental to energy balance and weight control \cite{WHO2020}. It also has important effects on the body, contributing to improved cardiovascular health, bone health, physical fitness, and mental well-being.

\subsubsection{Relationship of PA with cognitive readiness}
\label{pa}
PA has been shown to positively affect attention \cite{kramer2006exercise,vanhelst2016physical}. By monitoring PA, one can potentially gauge an individual's attention levels. Regular PA has been found to enhance cognitive function, including aspects like attention and memory \cite{hillman2008smart}. Someone engaged in light to moderate PA before a cognitive task might experience increased arousal, alertness, and readiness to perform the task \cite{tomporowski2003effects}. PA is found to promote resilience \cite{belcher2021roles,hegberg2015physical}. Studies suggest that PA can improve math performance \cite{have2018classroom,pindus2016moderate}. It can also significantly reduce reaction time \cite{brisswalter1997influence,abourezk1995effect} and increase task accuracy \cite{caponnetto2021effects,booth2013associations}. Further, it promotes reasoning \cite{perrot2009physical,wandzilak1988values}, reduces mental distress \cite{sciamanna2017physical} and helps in regulating emotion \cite{tang2022relationship,aparicio2016role}. 

\section{Neural Measurements}
\subsection{Brain Waves (BW)}
BW are patterns of electrical activity occurring in the brain. They are crucial to all aspects of brain functioning, such as thoughts, emotions, and behaviors. There are different types of BW, including alpha, beta, theta, delta, and gamma, each associated with different states of consciousness and cognitive activities \cite{yaomanee2012brain,duvinage2012p300}. 

\subsubsection{Relationship of Brain wave measures with cognitive readiness}
\label{bw}
Electroencephalogram (EEG) can be used to measure neural oscillations related to working memory, such as theta and alpha oscillations \cite{roux2014working,ruchkin1995working}. It can detect changes in neural activity patterns that may be associated with mental resilience, such as frontal asymmetry \cite{davidson2004does,uhlhaas2017youth}. BW can recognize the attention levels of individuals \cite{chen2017assessing,thomas2016eeg}. The latency of slow waves increases when doing difficult calculative tasks \cite{pauli1994brain,crk2015understanding}. EEG-derived measures like event-related potentials can be used to assess reaction time \cite{luck2005introduction}. Bartosz et al. found a relation between pilots’ concentration levels and reaction time, a positive correlation between theta power in the frontal lobe and response time \cite{binias2023analysis}. EEG-derived measures can also detect error-related negativity and error positivity, which are associated with error processing and accuracy \cite{falkenstein1991effects,craig2012regional}. Neural activity related to reasoning tasks, including changes in oscillatory activity, can also be measured using EEG \cite{bonnefond2012alpha,kao2012analysis}. It can also capture neural correlates of strategic decision-making, such as the P300 component \cite{sanfey2003neural,craig2012regional}. The BW activity of the frontal and central areas can give insight into positive (eustress) and negative (distress) stress \cite{flo2011transient}. BW like alpha, inhabited theta and high beta can give insight into relaxation and emotion regulation \cite{park2022effects,pollatos2012attenuated}.

\section{Metabolic Measurements}
\subsection{Hydration}
The process of providing adequate fluids to the body to ensure that all physiological processes can occur properly is hydration. Proper hydration is essential for body temperature regulation, joint lubrication, nutrient transportation, and various other critical bodily functions. 

\subsubsection{Relationship of Hydration with Cognitive Readiness}
\label{hydration}
Cognitive functions, such as attention, memory, and executive function, can be negatively affected by even mild dehydration \cite{adan2012cognitive,pross2017effects}. The detrimental effects of dehydration on cognitive performance include decreased concentration, alertness, and short-term memory, as well as increased feelings of fatigue \cite{masento2014effects}. Dehydration can also affect mood and increase the perception of task difficulty, which could further impact cognitive performance. Cognitive abilities such as memory, calculative work, reaction time, accuracy, and sustained attention were found to have negative effects as a result of decreased hydration level \cite{gopinathan1988role,cian2001effects,patel2007neuropsychological}. Reasoning was also affected by dehydration \cite{lieberman2007hydration}. With a decrease in water levels in the body, reasoning and decision-making capabilities, along with control over emotion, tend to decrease \cite{del2012hydration}.

\section{Compiling an Overview of Physiological Measures}
By comprehensively analyzing the relationship between physiological parameters and cognitive components, we can develop a deeper perspective of an individual's overall cognitive readiness. This, in turn, will lead to improved performance and outcomes in various cognitive tasks, like calculation, decision, reasoning, etc. However, most of these physiological measurements provide indirect information. This inherent indirectness makes it difficult for us to rely solely on one specific type of physiological measure when attempting to analyze an individual's cognitive readiness, leading researchers to overlook other critical aspects. To approach this challenge more effectively, it is preferable to compile a catalog of cognitive readiness components that consider as many physiological measures as possible. In doing so, we can obtain a more holistic and accurate understanding of cognitive readiness. 
The catalog concept has already gained significant attention and proven its importance in various other fields of research \cite{van1944perspective,salaba2023cataloging,joachim2003historical}. Therefore, we propose a flow diagram and discussion of a catalog system for cognitive readiness in Section 13.

Tables \ref{table1} and \ref{table2} represent a list of different physiological measures and their contribution to our understanding of cognitive readiness. This can be considered as a basic building block for a catalog to assess cognitive readiness with these measures. The tables clearly demonstrate that measures such as HR, HRV, sleep, PA, and BW can provide insights into multiple aspects of cognitive readiness. Conversely, measures like BFR \& BOL, BR, PD, EB and hydration might offer an understanding of fewer cognitive readiness components but still play a crucial role in comprehending an individual's overall cognitive readiness. However, a limitation to these tables is a lack of description of the strength between the physiological measures and the domains of cognitive readiness.

\subsection{Wearables in measuring physiological parameters}
While physiological measures provide a crucial understanding of an individual's cognitive readiness, the challenge lies in the practical, real-time acquisition of this data. The integration of technology into our everyday lives has led to significant transformations, creating a paradigm where machines and humans co-exist and interact closely. The evolution of HCI tools, in particular, has paved the way for real-time data capture of various physiological and cognitive markers \cite{brown1999human}, providing profound insights into an individual's cognitive readiness. This concept has led to a surge in the demand for HCI tools, given their potential to significantly enhance performance across diverse sectors. Amid this burgeoning interest, wearable devices have emerged as powerful HCI tools, thanks to their ability to capture and analyze physiological measures in a non-invasive, convenient, and real-time manner \cite{bowman2021wearable,peake2018critical}. Tracing the development of wearable devices back to the invention of the pocket watch in the 17th century, the progress in this space has been phenomenal \cite{historyWearable}. The late 20th century saw the rise of first-generation wearable devices like digital watches and portable calculators \cite{mann1997wearable}. Then, in the late 1980s and early 1990s, Steve Mann pioneered wearable computing, sparking a new wave of innovation. This was followed by the advent of fitness trackers in the 2000s, with devices like the \textit{Fitbit} that leveraged sensor technology to monitor health-related parameters \cite{kim2019wearable}. The trend escalated in the 2010s when major tech players like Apple and Google introduced smartwatches and augmented reality glasses \cite{lv2015touch}. The \textit{Apple Watch}, in particular, set a new benchmark for wearable tech with its advanced features. 

As of the time of this paper, wearable technology is heading towards more specialized wearable devices with advanced bio-sensing features for health monitoring. Innovations like smart rings, smartwatches, wristbands, EEG headbands, biomechanical shoes, smart garments, wearable skin patches, inertial measurement units (IMUs), smart glasses, chest straps, and wearable pulse oximeters are on the rise. These devices can track a variety of physiological measures. This, in turn, enables real-time, in-the-wild assessment and feedback, which is crucial for optimizing cognitive performance and overall mental well-being.

\begin{figure*}[!t]
    \centering
    \includegraphics[width=.9\textwidth]{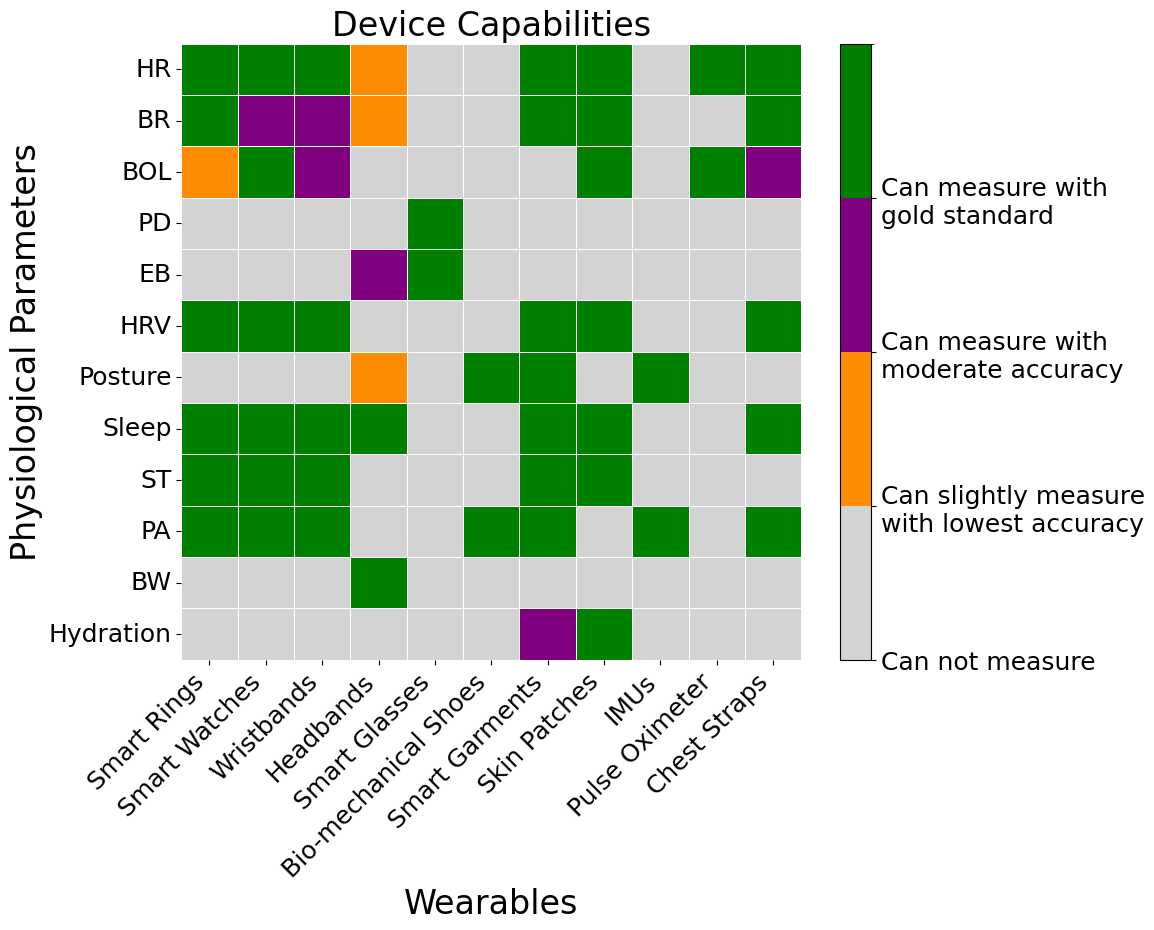}
    \caption{{Pictorial presentation of wearables' standards to access different physiological measures}}
    \label{fig2}
\end{figure*}

Therefore, we have constructed a heatmap, shown in Fig:\ref{fig2}, by surveying the reports \cite{stone2021assessing,fiorini2021characterization,e2023validation,niela2022comparison,mehrabadi2020sleep,theurl2023smartwatch,sarhaddi2022comprehensive, hernandez2015biowatch,jiang2023investigating,xie2018evaluating,arnal2019dreem,lee2023variations,chang2016detection,thorey20201211,lee2023preliminary,zhang2017using,angelucci2021smart,kim2021smart,verdel2022validity,ardalan2020towards,singh2017comparative,cosoli2020wrist} to find which wearable is well-suited for accessing a particular physiological measure. The {horizontal axis} shows different types of wearable devices, and the {vertical} axis is marked with different physiological measures. Each cell in the heatmap represents the capability of a specific wearable device to track a particular physiological measure. In the heatmap we see different standards of accuracy (gold, moderate, lowest) of wearables to physiological measure. The light gray color indicates that the device can not measure the indicated physiological parameter. The orange color represents that the device is capable of measuring with the lowest level of trustable accuracy, while the purple color represents that the device is partially capable of measuring the physiological measure with moderate accuracy. Finally, the green block indicates that the device can measure the indicated physiological measure with a gold standard level of accuracy. One can easily observe from this figure that smart watches, smart garments, smart rings, wristbands \cite{pitzalis2023state}, and chest straps are the most capable devices in measuring different physiological measures with either a gold standard or an acceptable range of accuracy. On the other hand, devices like EEG headbands, bio-mechanical shoes, IMUs, and smart glasses might display a sparser pattern, with green appearing only under very specific physiological measures. This reflects their specialized tracking capabilities. This figure emphasizes the notion that no single device is universally optimal for tracking all health metrics, underscoring the need for the purposeful selection of wearable technology based on the specific physiological metrics of interest. 

While the heatmap provides an informative general overview, it is important to remember that the exact specifications and capabilities of wearable devices can vary significantly based on model and manufacturer. Therefore, the heatmap should be regarded as a generalized guideline.

\section{Techniques used in wearable devices}
Wearable devices employ a variety of advanced techniques and technologies to measure the physiological parameters discussed above. Several different sensors are used for this purpose. The most common methods are discussed in this section.

\subsection{Photoplethysmography (PPG)}
This technique is used by most fitness trackers, smartwatches, and smart rings to measure HR and HRV, monitor sleep (via HR monitoring), infer BR (using light to measure the volume of an organ), and measure BFR (by measuring blood volume changes in capillaries). It involves shining a light onto the skin and measuring the amount of light that is scattered by blood flow, which changes with the pulse \cite{allen2007photoplethysmography,namvari2022photoplethysmography}. In simpler terms, PPG detects blood volume changes in the microvascular bed of tissue used to monitor a patient's heartbeat. A variety of light sources (LEDs) and light detectors can be used for this, and they are usually placed in a configuration that allows the light to pass through or reflect off the skin and into the detector. The device measures the light absorption of blood to calculate HR and inter-beat intervals to calculate the HRV. When the heart pumps, the blood flows, and thus the green light absorption is greater. By flashing the LED lights hundreds of times per second, devices can calculate the number of times the heart beats per minute (bpm) \cite{shi2013principles}.

\subsection{Electrocardiography (ECG/EKG)} 
This technique is used by some more specialized wearable devices. ECG is a well-established measurement technique that uses electrodes placed on the skin to measure the electrical activity of the heart. Therefore, it is used for monitoring HR, HRV, and sleep \cite{nishad2018application}. The electrodes detect tiny electrical changes on the skin that arise from the heart muscle depolarizing during each heartbeat \cite{laguna1997database}. In a standard ECG test, ten electrodes are placed on the patient's limbs and on the surface of the chest. The signals are then amplified and recorded. Wearable ECG devices, on the other hand, often use fewer electrodes and may need to be in contact with different parts of the body simultaneously (like the chest and wrist) in order to get a reading. A well-known example of a wearable with EKG capabilities is the \textit{Apple Watch} Series 4 and up \cite{isakadze2020useful}, which have electrical heart sensor pairs in the Digital Crown and on the back crystal.

\subsection{Accelerometry} 
Accelerometry is performed by integrating accelerometers into fitness trackers, smartwatches, or specialized research-grade wearable devices to track movements \cite{freedson1998calibration}. Therefore, they can give insight into sleep (based on actigraphy)  \cite{sadeh2011role}, BR (by measuring the rise and fall of the user's chest or abdomen during breathing), PA \cite{matthews2012best}, posture and body positioning \cite{chen2005technology}. These devices can collect continuous or intermittent acceleration data, which is then processed and analyzed to extract the desired physiological measures.

\subsection{Oximetry} 
In oximetry, two light beams are shined into the small blood vessels or capillaries of a person's finger, earlobe, or other tissue \cite{nitzan2014pulse}. These beams of light are of two different wavelengths: one is red light and the other is infrared light \cite{wilson2010accuracy}. The ratio of the amount of red light absorbed, compared to the amount of infrared light absorbed, is calculated by the sensor, giving the oxygen saturation level (BOL) as a percentage \cite{southall1987pulse}. The absorption of light changes with the pulsing (expansion and contraction) of the arterial blood vessels, so the sensor can also measure HR.

\subsection{Temperature Sensing} 
Most wearable devices use one of two types of temperature sensing techniques to measure the ST of the wearer. The first type is resistance temperature detectors (RTDs) \cite{childs1999tympanic,li2016wearable}. These sensors measure temperature by correlating the resistance of the RTD element with body temperature. As the temperature changes, the electrical resistance of the material also changes. The most commonly used RTD element is made of platinum. The second type is thermistors. Thermistors are a kind of resistor whose resistance varies significantly with temperature. They are made from semiconductor materials. Negative temperature coefficient (NTC) thermistors are widely used in wearables because of their high sensitivity, small size, and low cost \cite{tamura2018current}.

\subsection{EEG} 
EEG-based wearables can measure the electrical activity of BW by placing sensors on the scalp \cite{klimesch1999eeg}. These activities are then interpreted and correlated with cognitive work.

\subsection{Electrooctoculography (EOG)} 
EOG is a method of measuring the electrical activity of the muscles that control eye movement. It can be used to measure eye movement, eye blinks, and other eye-related phenomena \cite{deng2010eog}. EOG can be used to assess the cognitive readiness of a person by measuring their attention, focus, and alertness. Borghini et al. \cite{borghini2016quantitative} showed that EOG can be used to predict task performance in a variety of tasks, including reading, memory, and problem-solving. It can also be used to detect fatigue and drowsiness, \cite{zhu2014eog} which, in turn, can be used to assess a person's cognitive readiness while performing tasks that require high levels of attention and focus. Tag et al. \cite{9115246} found using EOG that fatigued subjects have more frequent EBs than their non-fatigued counterparts.

\subsection{Bioimpedance Analysis (BIA)} 
BIA is a method used to estimate body composition, particularly body fat and muscle mass. It is based on the principle that different types of body tissues (fat, muscle, bone, etc.) have different electrical conductivities. In wearable devices, BIA is performed by sending a small, unnoticeable electric current into the body from one part of the device and then measuring the voltage drop at a different point on the device \cite{lee2015comparison}. The device then uses this voltage drop to calculate the impedance. This is used to analyze the hydration level of the user \cite{rosler2010nutritional}.

\section{Prospects} 
Following our comprehensive analysis of the links between physiological measures and cognitive components, we suggest the development of a catalog that offers a broad view of cognitive readiness. This catalog would provide a wide-ranging understanding of cognitive readiness that applies to virtually everyone, regardless of their specific situation. This requires that the catalog be easily and quickly updated and annotated to include new forms of measurements and account for different levels of relationships between measurements and domains. By including a plethora of physiological measures, we ensure that all crucial factors of cognitive readiness are considered. However, it is also true that different situations might require focusing on specific aspects of cognitive readiness. For example, a student in a classroom might need to concentrate more on attention and memory, while a pilot might need to focus on quick decision-making and reaction times. This situation-specificity has to be considered by all designers, researchers, and developers concerned with cognitive readiness. 

\begin{figure*}[!t]
    \centering
    \includegraphics[width=1\linewidth]{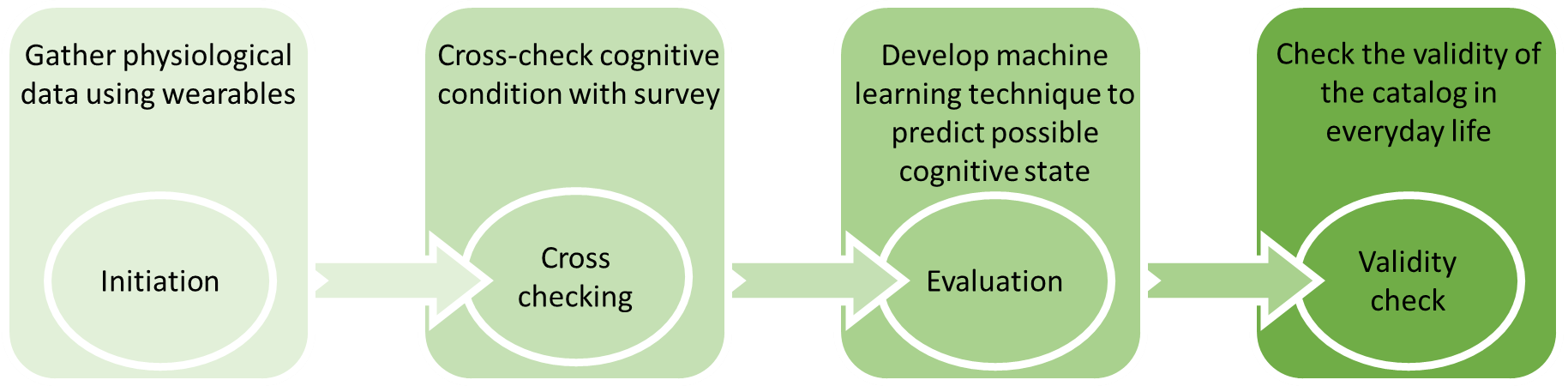}
    \caption{{Conceptual flow diagram for a situational catalog.}}
    \label{fig:2}
\end{figure*}

Therefore, our proposed catalog would be built to provide a foundation that can be tailored to these specific needs. We can easily create more specific guides in response to the needs of particular situations. These situational catalogs are more precise, matching the right physiological measures to the most important cognitive skills needed in those situations. This makes our catalog of cognitive readiness even more useful and applicable in real-world settings. In Figure \ref{fig:2}, we propose a flow diagram to create a catalog for a particular situation for diverse applications.

Researchers can use the appropriate wearables to collect real-time physiological data. This data is then transmitted to a large database. The goal is to effectively map out how physiological measures correlate with different cognitive conditions for a particular situation. A well-developed machine learning model allows the system to recognize patterns in the physiological data and predict an individual's cognitive readiness. This predictive capability is key to developing personalized brain training programs. These programs can offer specific exercises and interventions tailored to an individual's needs. As individuals engage with the personalized training, their responses and improvements would be fed back into the database. This continuous cycle of data collection, analysis, and feedback ensures that the system keeps evolving, becoming more accurate in its predictions and effective in its training suggestions. Over time, this would enable people to maximize their cognitive potential, receiving support and training that is perfectly aligned with their unique physiological makeup. We see several significant implications of this proposed catalog:

One of the foremost advantages of creating a cognitive readiness catalog is its potential for personalization. With an increasing emphasis on personalized healthcare and learning, a catalog that maps specific physiological measures to levels of cognitive readiness can be instrumental. It would allow for interventions, be it cognitive training or stress management techniques, to be customized to the individual, thereby increasing their effectiveness.
The predictive capabilities of such a catalog cannot be overstated. By tracking an individual's physiological measures and referencing the cognitive readiness catalog, we could anticipate their state of cognitive readiness. This ability to forecast could be particularly useful in high-stress or high-stake scenarios, where readiness to respond is critical, thereby preventing mishaps and enhancing overall performance.
The catalog would also serve as a standard for cognitive readiness. It would allow for comparison across different populations or fields, fostering a more nuanced understanding of cognitive readiness. This benchmarking ability is vital to areas such as education, sports, or professional settings where an individual's readiness directly influences their performance and outcomes.
The catalog of cognitive readiness would be a valuable resource for research, presenting a systematic and comprehensive way to understand the interplay between physiological parameters and cognitive readiness. It could lead to the discovery of new correlations, thereby further advancing the field.
By giving individuals, educators, employers, and healthcare providers a clearer picture of cognitive readiness, the catalog can facilitate better-informed decision-making. It can guide educational strategies, work schedules, or even lifestyle changes to optimize cognitive performance and overall well-being. 

With continued advancements in wearable technology and cognitive science, we are optimistic about the transformative impact of this endeavor, that is, the creation of a cognitive readiness catalog based on physiological measures, on the way we understand and optimize cognitive performance.

\section{Conclusion}
In this review, we find that interpreting cognitive readiness in real-world scenarios is complex due to diverse physiological measures, individual variability, and limitations of wearable devices. Considering an in-depth assessment of cognitive readiness for particular situations, we believe a catalog system of key physiological measures can serve as proxies for a detailed assessment of cognitive readiness. Therefore, the catalog can be an excellent tool to assess cognitive readiness for diverse applications, with special focus on in-the-wild research settings.

\section*{Acknowledgments}
This work was supported in part by JST Trilateral AI Research (JPMJCR20G3), JSPS Grant-in-Aid for Scientific Research (23KK0188), and the Grand Challenge of the Initiative for Life Design Innovation (iLDi).

\bibliography{bibliography}
\bibliographystyle{ACM-Reference-Format}

\end{document}